\documentclass[aip,jmp,amsmath,amssymb,preprint]{revtex4-1}

\usepackage{graphicx}% Include figure files
\usepackage{dcolumn}% Align table columns on decimal point
\usepackage{bm}% bold math
\begin{document}

\title{Two-loop Feynman integrals for $\phi^4$ theory with long-range correlated disorder}

\author{M. Dudka}

\affiliation{
Institute for Condensed Matter Physics, National
Academy of Sciences of Ukraine, UA--79011 Lviv, Ukraine
}
\date{Lviv, \today}
\begin{abstract}
Two-loop massive Feynman integrals for $\phi^4$ field-theoretical model with long-range correlated disorder are considered. Massive integrals for the vertex function $\Gamma^{(4)}$ including two or three massless propagators  for generic space dimension  and for any value of the correlation parameter  are evaluated analytically applying Mellin-Barnes method as well as familiar representation for one-loop integrals.
 % using known results. 
Obtained expressions are presented in the form of hypergeometric functions.
\end{abstract}

\maketitle

\section{Introduction}
Feynman integrals are known to appear in various branches of theoretical physics exploiting perturbative quantum field theories. Recent interest  to these objects in high energy physics is generated by necessity of evaluation of radiative corrections (expressed in terms of complicated Feynman integrals)  to compare analytical results with experimental data of Large Hadron Collider. To achieve satisfactory accuracy it requires exploitation of high-loop order integrals. However even  evaluation of certain low-order  Feynman integrals is not a trivial task since there is no unique receipt   for calculation of every Feynman integral.  Rather Feynman integrals can be successfully evaluated combining different methods \cite{Smirnov}.

    Feynman integrals are frequently used also in statistical physics, where the quantum field theory is exploited for the critical phenomena description. There Feynman integrals are involved in different renormalization group (RG) schemes \cite{rgbooks} for calculation of physical observables, like critical exponents, amplitude ratios, scaling functions, {\em etc}. Being universal quantities, these observables   depend only on global characteristics of a system, like dimension of space $d$ , order parameter dimension $N$, internal symmetries. One of the fundamental problems there is dependence of the results for general space dimension $d$.

     %There dependence of results for general space dimension $d$ presents special interest, since it allows to check their correctnes or to obtain some %new insights.

    Usually such dependence can be studied within famous Wilson-Fisher expansion \cite{WilsonFisher1972}
    in deviation $\epsilon$ of  a space dimension from its upper critical value. $O(N)$ symmetric $\phi^4$ theory has 4 as an upper critical dimension  and thus $\epsilon=4-d$.  High orders of the $\epsilon$-expansion allow to obtain accurate values for critical exponents of three-dimensional $O(N)$ models \cite{Guida98}.  Calculation of critical exponents within massive field theory directly at  fixed space dimension $d=2$ or $d=3$ proposed by Parisi \cite{Parisi} serves as an alternative to this method. To apply such approach one needs to know the values of massive Feynman integrals  for given $d$ within considered loop number. Estimates of critical exponents on the base of this method  for the three-dimensional $\phi^4$ theory are in correspondence with the values given by $\epsilon$-expansion \cite{Guida98}. This method was successfully used to  study   modifications of $\phi^4$ model for three-dimensional systems  with cubic anisotropies \cite{CarmonaPelissettoVicari2000,FolkHolovatchYavorskii2000,PelissettoVicari}, uncorrelated weak quenched disorder \cite{GrinsteinLuther,Shpot89,MayerSokolovShalaev,PakhninSokolov,PelissettoVicari2000,PelissettoVicari,HolovatchDudkaYavorskii,Holovatch02,FolkHolovatchYavorskii2003}, multicriticality \cite{PrudnikovPrudnikovFedorenko98}, random anisotropies \cite{Dudka01b,Dudka01a,DudkaHolovatchFolk01,Holovatch02,Calabrese04}, frustrations \cite{PelissettoRossiVicari,PelissettoVicari,DelamotteDudkaHolovatchMouhanna,DelamotteDudkaHolovatchMouhannaa}. Note that these modifications do not change structure of massive Feynman integrals for the $\phi^4$ theory.

    Fixed space dimension approach can be also  used for studies at non-integer values of $d$.
    Thus  numerical calculation of massive integrals  of $\phi^4$ theory for general dimension $2<d<4$  was performed within two-loop order in a study of disordered Ising system \cite{HolovatchShpot}. The numerical values for three-loop massive integrals were later obtained for $0<d<4$ \cite{HolovatchKrokhmalskii} and used for analysis of critical properties in random Ising in space dimension range $2<d<4$ \cite{HolovatchYavorskii}. Recently dependence of two-loop massive integrals of $O(N)$ symmetric $\phi^4$ field theory was calculated analytically obtaining integrals in a compact form of Gauss hypergeometric functions with $d$-dependent parameters \cite{Shpot09}.

      However field-theoretical description of some complex  systems may include also modified integrals.
      For instance, introduction of an anisotropy to interactions  as in a system with Lifshitz critical point \cite{Hornreich1975} results  in Feynman integrals with different masses in propagators. Recent  calculations of corresponding one-loop massive integral    for general  $d$ present  results in form of hypergeometric Appell functions  as well as new reduction for them \cite{Shpot07}. Anisotropy  in correlation also appears for models of magnets with non-magnetic defects that are correlated in $\epsilon_d$ dimension and randomly distributed in $d-\epsilon_d$ space \cite{Dorogovtsev,BoyanovskyCardy82,LawriePrudnikov84,Blavatskaa,Blavatskab,Blavatskac}. Presence of such defects also changes the structure of Feynman integrals.

      Here we are interested in another kind of correlated disorder, where correlations between defects decay at large distance $R$ between them according to power law $R^{-a}$.   As it is shown below, massive Feynman integrals in  this case may have additional massless propagators.  Effects of such  long-range correlated disorder  on the critical properties of $\phi^4$ model were studied intensively  including static critical behaviour \cite{WeinribHalperin83,KoruchevaUzunov84,PrudnikovFedorenko99,PrudnikovPrudnikovFedorenko99,PrudnikovPrudnikovFedorenko00}, critical dynamics near equilibrium \cite{PrudnikovFedorenko99,PrudnikovPrudnikovFedorenko99,PrudnikovPrudnikovFedorenko00,KoruchevaDeLaRubia98}, short time  critical dynamics \cite{ChenGuoLi01,LiChenGuo01,ChenLi05}, critical ultrasound propagation \cite{PrudnikovPrudnikov09}.
      Disorder with long-range correlations appears in systems of different nature. Studies  the phase transition in superconductor    with long-range correlated impurities\cite{KoruchevaMillev94},   statics and dynamics of elastic systems in disordered media \cite{FedorenkoLeDoussalWiese06,Fedorenko08}, conformal properties of polymers in disordered environment \cite{Holovatch02,Blavatska_vonFerberHolovatch01}, quantum critical behaviour in systems with long-range correlated impurities \cite{Takov01}, percolation in correlated systems \cite{Weinrib84,Marinov06} can be mentioned. Recently two-dimensional fermionic system with long-range correlated disorder  relevant for description of disordered graphene  was investigated  \cite{FedorenkoCarpentierOrignac12}.

In this paper, analytical calculations of two-loop massive Feynman integrals for systems with long range correlated disorder are performed for general space dimension $d$ and correlation parameter $a$. Such integrals were known analytically  only in one-loop order for $d=3$ and general $a$, in two-loop order they were calculated numerically for $d=3$ and $2<a<3$ \cite{PrudnikovPrudnikovFedorenko99,PrudnikovPrudnikovFedorenko00}. A challenge of this paper is to obtain expressions for two-loop integrals for general space dimension $d$ and correlation parameter $a$ via known functions.
The set-up of the paper is the following. In Section \ref{II}   field theoretical description for systems with correlated disorder  as well as  two-loop integrals appearing within such description are presented. In the next Section \ref{III} the one loop expressions are considered integration over internal momentum was performed. In the Section \ref{IV}   results for two-loop integrals  are presented in form of hypergeometric functions with $d$- and $a$- dependent parameters. Section \ref{V} summarizes the paper.  Definitions of functions as well as some intermediate calculations are given in the Appendices.

\section{Field-theoretical model with correlated disorder}\label{II}

We consider  system with quenched defects having correlation function $g(R)$ dependent on the distance $R$ between them. To deal with quenched disorder one should average free energy  over disorder configurations. It can be performed with help of replica trick, that gives  effective Hamiltonian of $\phi^4$ type \cite{WeinribHalperin83}:
\begin{equation}\label{Hamiltonian}
\mathcal{H}=\sum_{\alpha=1}^n\int d^d R\left[ \frac{1}{2}\left(r_0\phi^2_\alpha+(\nabla\phi_\alpha)^2+\frac{u_0}{4!}(\phi^2_\alpha)^2\right)\right]
-\sum_{\alpha,\beta=1}^n\int d^d R d^dR'g(|R-R'|)\phi^2_\alpha(R)\phi^2_\beta(R')
\end{equation}
 Here, $\phi_\alpha$ is $N$-component vector. In this field-theoretical model parameter $r_0$ is  a linear function in temperature and plays a role of a bare mass, $u_0>0$ corresponds to bare coupling of $O(N)$-symmetrical model.
The long-distance properties of (\ref{Hamiltonian}) in the replica limit $n\to 0$  describe critical behaviour occurring in the disordered system involving  correlations between defects.  In the model considered here, disorder correlations weaken according  to the power law $g(R)\sim R^{-a}$ for large separation $R$ \cite{WeinribHalperin83}. Furier transform of the correlation function for defects gives
\begin{equation}\label{furier}
{\bar g}({\bf k})=v_0+w_0 k^{a-d}
\end{equation}
 for small $k$. In the case  $a>d$ the second term of ${\bar g}({\bf k})$  is irrelevant for $k\to0$ and therefore this case corresponds to short range disorder. We are interested in the case when $a<d$ that makes term with $w$ coupling  crucial at small $k$. On of he interpretation of the model is that disorder correlation function with $a=d-1$ corresponds to the case of straight lines of impurities with random orientation, while case $a=d-2$ corresponds to random planes of impurities.

 Standard tool to describe critical behavior is application of field theoretical renormalization group approach \cite{rgbooks}. In this approach, the vertex functions $\Gamma^{(n)}$ are considered, their finiteness is ensured by imposing certain normalisation conditions. Using the renormalization at fixed mass and zero external momenta \cite{Parisi}  one has to calculate massive Feynman integrals involving momentum integration of dimension $d$. Feynman integrals for the Hamiltonian (\ref{Hamiltonian}) with (\ref{furier}) appearing within two-loop approximation were presented  in form of Feynman diagrams together with diagrammatic rules in \cite{PrudnikovPrudnikovFedorenko00}.  In general case these integrals  can be written as:
\begin{equation}\label{integrals}
I(\alpha,\beta,\gamma,\delta,\rho,\tau)=\int_{q_1}\int_{q_2}
\frac{(q_1^{\alpha}q_2^{\beta}|{\bf q}_1+{\bf q}_2|^{\gamma})^{a-d}}{(r+q_1^2)^\delta(r+q_2^2)^\rho(r+({\bf q}_1+{\bf q}_2)^2)^\tau},
\end{equation}
where we omit in the notation dependence on $r$, $a$ and $d$ for simplicity. Integration in (\ref{integrals}) means:
\begin{eqnarray}\label{spher}
\int_q\{{\ldots}\}&{=}&\frac{1}{(2\pi)^d }\int d{\bf q}\{\ldots\}=\frac{1}{(2\pi)^d }\prod_{i=1}^{d{-}2}\int_0^{\pi}d\theta_i \sin^i\theta_i\int_0^{2\pi}d\varphi\int_0^{\infty}q^{d{-}1}dq\{\ldots\}.
\end{eqnarray} Numerators in (\ref{integrals}) appear for diagrams including renormalized coupling $w$  only. At $a=d$ we get in (\ref{integrals})  a form of two-loop integrals relevant to usual $\phi^4$ theory \cite{rgbooks}.

%Diagramatic rule for this $w$-vertex in Furier space is given in Fig.~\ref{fig1}.
% \begin{figure}[htbp]
%{\includegraphics[width=0.4\textwidth]{diagramm2.eps}}
%\caption{\label{fig1} Diagrammatic rule for $w$-term of Hamiltonian (\ref{Hamiltonian}) carring the
%additional momentum dependence.}
%\end{figure}

Calculations of Feynman diagrams with one $w$-vertex (with one momentum in the numerator on (\ref{integrals})) can be simply performed by standard integration methods, while evaluation of    Feynman integrals with two or three $w$-vertices presents more difficult task. Here, calculating four-point vertex function $\Gamma^{(4)}(k_i,r,{u_i})$ %, which consists in three parts $\Gamma^{(4)}_u$, $\Gamma^{(4)}_v$, $\Gamma^{(4)}_w$
one meets nine such integrals:
 \begin{eqnarray}\label{int1}
 I(1,1,0,2,0,2)&=&\int_{q_1}\int_{q_2}
\frac{q_1^{(a-d)}q_2^{(a-d)}}{(r+q_1^2)^2(r+({\bf q}_1+{\bf q}_2)^2)^2}, \\ \label{int2}
 I(1,1,0,3,0,1)&=&\int_{q_1}\int_{q_2}
\frac{q_1^{(a-d)}q_2^{(a-d)}}{(r+q_1^2)^3(r+({\bf q}_1+{\bf q}_2)^2)},\\ \label{int3}
  I(1,1,0,2,1,1)&=&\int_{q_1}\int_{q_2}
\frac{q_1^{(a-d)}q_2^{(a-d)}}{(r+q_1^2)^2(r+q_2^2)(r+({\bf q}_1+{\bf q}_2)^2)},\\ \label{int4}
  I(1,1,1,2,0,2)&=&\int_{q_1}\int_{q_2}
\frac{q_1^{(a-d)}q_2^{(a-d)}|{\bf q}_1+{\bf q}_2|^{(a-d)}}{(r+q_1^2)^2(r+({\bf q}_1+{\bf q}_2)^2)^2},\\ \label{int5}
  I(2,1,0,3,0,1)&=&\int_{q_1}\int_{q_2}
\frac{{q_1^2}^{(a-d)}q_2^{(a-d)}}{(r+q_1^2)^3(r+({\bf q}_1+{\bf q}_2)^2)},\\ \label{int6}
 I(2,1,0,2,1,1)&=&\int_{q_1}\int_{q_2}
\frac{{q_1^2}^{(a-d)}q_2^{(a-d)}}{(r+q_1^2)^2(r+q_2^2)(r+({\bf q}_1+{\bf q}_2)^2)},\\ \label{int7}
  I(1,2,0,2,1,1)&=&\int_{q_1}\int_{q_2}
\frac{q_1^{(a-d)}{q_2^2}^{(a-d)}}{(r+q_1^2)^2(r+q_2^2)(r+({\bf q}_1+{\bf q}_2)^2)},\\ \label{int8}
  I(1,1,0,1,1,2)&=&\int_{q_1}\int_{q_2}
\frac{q_1^{(a-d)}q_2^{(a-d)}}{(r+q_1^2)(r+q_2^2)(r+({\bf q}_1+{\bf q}_2)^2)^2},\\
  \label{int9} I(1,1,1,2,1,1)&=&\int_{q_1}\int_{q_2}
\frac{q_1^{(a-d)}q_2^{(a-d)}|{\bf q}_1+{\bf q}_2|^{(a-d)}}{(r+q_1^2)^2(r+q_2^2)(r+({\bf q}_1+{\bf q}_2)^2)}.
  \end{eqnarray}
Main goal of this paper is evaluation of these nine integrals. It is done in the next two sections.

\section{Evaluation of internal integrals}\label{III}

To  calculate integrals (\ref{int1})-(\ref{int9}) presented above we perform subsequent integration over two internal momenta.
Let us first perform an integration over $q_2$.  As one can see the integrals over $q_2$ for (\ref{int1})-(\ref{int9}) can be represented as
four $q_1$ -dependent functions:
\begin{eqnarray}\label{fan1}
f_1(\alpha,\beta,q_1)&=&\int_{q_2}\frac{1}{(q^2_2)^\alpha(r+({\bf q}_1+{\bf q}_2)^2)^\beta},\\
\label{fan2}
f_2(\alpha,\beta, q_1)&=&\int_{q_2}\frac{1}{({q}_2^2)^\alpha(({\bf q}_1+{\bf q}_2)^2)^\alpha(r+{ q}_2^2)^\beta},
\\
\label{fan3}
f_3(\alpha,\beta,\gamma,q_1)&=&\int_{q_2}\frac{1}{(q_2^2)^\alpha(r+q_2^2)^\beta(r+({\bf q}_1+{\bf q}_2)^2)^\gamma},
\\
\label{fan4}
f_4(\alpha,\beta,\gamma,q_1)&=&\int_{q_2}\frac{1}{(q_2^2)^\alpha(({\bf q}_1+{\bf q}_2)^2)^\alpha(r+q_2^2)^\beta(r+({\bf q}_1+{\bf q}_2)^2)^\gamma}.
\end{eqnarray}
In (\ref{fan1})-(\ref{fan4}) values of integer parameters $\gamma$ and $\beta$ equal to 1 or 2, while parameter  $\alpha=(d-a)/2$ or $\alpha=(d-a)$.
We are interested in the case  $a<d$, for which correlations of defects are relevant. Then the value of $\alpha$   is positive, that justifys form of (\ref{fan1})-(\ref{fan4}).  Therefore we have integrals with massive and massless propagators.

To obtain expressions for above functions (\ref{fan1})-(\ref{fan4}) we  appeal to  the method
of evaluating massive Feynman integrals  based on the representation of massive denominators in the form of the Mellin-Barnes
contour integrals. This method was  developed in Refs. \cite{BoosDavydychev,Davydychev1991,Davydychev1992} for evaluation of similar integrals.

First integral $f_1$ can be simply obtained by several method. We present its calculation along the lines of Ref.  \cite{BoosDavydychev}  in the Appendix~\ref{ApB}. The result reads:
\begin{equation}\label{f1}
f_1(\alpha,\beta,\vec{q}_1)=(r)^{\frac{d}{2}-\alpha-\beta}
\frac{S_d \Gamma(\alpha+\beta-\frac{d}{2}) \Gamma(\frac{d}{2}-\alpha)}{2\Gamma(\beta)}
{_2F_1\left[\left.\begin{array}{c}
\alpha,\alpha+\beta-\frac{d}{2}\\ \frac{d}{2}\end{array}\right|-\frac{q^2_1}{r}\right]},
\end{equation}
where $\Gamma(x)$ is the Gamma function,   $S_d=\frac{1}{2^{d-1} \pi^{d/2}\Gamma(d/2)}$, $_2F_1$ is the Gauss hypergeometric function \cite{Erdeley}. For definition of $_2F_1$ and its integral representation see (\ref{2f1}) and (\ref{2f1int}) in Appendix~\ref{ApA}.
Note, that results for usual $\phi^4$ theory are obtained as a particular case of our results at $a=d$ ($\alpha=0$ in (\ref{f1})). At $\alpha=0$ function $_2F_1$ in  (\ref{f1}) becomes unity and we come to known result (see e.g. \cite{rgbooks})

The obtained result  can be used for calculation of $f_2(\alpha,\beta,q_1)$. To do this we transform (\ref{fan2}) to the form similar to $f_1(\alpha,\beta,q_2)$ with the help of Feynman parametrisation (see \ref{fenpam}).
\begin{equation}\label{fey_par}
f_2(\alpha,\beta,q_1)=\frac{\Gamma(\beta+\alpha)}{\Gamma(\alpha)\Gamma(\beta)}\int_0^1 dx x^{\beta-1}(1-x)^{\alpha-1}\int_{q_2}\frac{1}{(q_2^2)^{\alpha}(rx+({\bf q}_1+{\bf q}_2)^2)^{\beta+\alpha}}.
\end{equation}
Substituting  result  (\ref{f1}) into  (\ref{fey_par}) instead the last integral we have
\begin{eqnarray}
f_2(\alpha,\beta,q_1)&=&\frac{S_d \Gamma(\beta{+}2\alpha{-}d/2) \Gamma(d/2-\alpha)}{2\Gamma(\alpha)\Gamma(\beta)}\times\nonumber\\&&\int_0^1 dx\, x^{\beta{-}1}(1{-}x)^{\alpha{-}1}(rx)^{d/2-2\alpha{-}\beta} {_2F_1\left[\left.\begin{array}{c}
{}\alpha,\beta{+}2\alpha{-}d/2\\ d/2\end{array}\right|{-}\frac{q^2_1}{rx}\right]}.
\end{eqnarray}

To calculate  the integral  we first  use the Mellin-Barnes representation for the function $_2F_1$ (see (\ref{2f1cont}) in Appendix~\ref{ApA}). Then, performing an integration over $x$ we get
\begin{eqnarray}
f_2(\alpha,\beta,q_1)&{=}&r^{d/2{-}2\alpha{-}\beta}\frac{S_d \Gamma(d/2)\Gamma(d/2{-}\alpha)}{2\Gamma({}\alpha)\Gamma(\beta)}\times\nonumber\\&&
\frac{1}{2\pi i}\int_{{-}i\infty}^{i\infty}ds\,
\frac{\Gamma({-}s)\Gamma({}\alpha{+}s)\Gamma(d/2{-}2\alpha{-}s)\Gamma({}2\alpha{+}\beta{+}s{-}d/2)}{\Gamma(d/2{-}\alpha{-}s)\Gamma(s{+}d/2)}\left(\frac{q^2_1}{r}\right)^s\!\!.
\end{eqnarray}
Considering the integral over $s$, one can see that in the right half-plane of the complex
variable $s$ there are two series of poles due to  $\Gamma({-}s)$ and $\Gamma(d/2{-}2\alpha{-}s)$. Therefore integration can be performed with the help of residue theorem and the result is the  following:
\begin{eqnarray}
f_2(\alpha,\beta,q_1)&=&\frac{r^{d/2{-}2\alpha{-}\beta}S_d \Gamma(d/2)\Gamma(d/2{-}\alpha)}{2 \Gamma(\beta)\Gamma({}\alpha)}\times\nonumber\\&&\left\{
\sum_{n\ge 0}\left({-}\frac{q^2_1}{r}\right)^n\frac{1}{n!}
\frac{\Gamma({}\alpha{+}n)\Gamma(d/2{-}2\alpha{-}n)\Gamma({}2\alpha{+}\beta{+}n{-}d/2)}{\Gamma(d/2{-}\alpha{-}n)\Gamma(n{+}d/2)}\right .{+}
\nonumber\\&&\left.\left(\frac{q^2_1}{r}\right)^{d/2{-}2\alpha}\sum_{n\ge 0}\left({-}\frac{q^2_1}{r}\right)^n\frac{1}{n!}
\frac{\Gamma(\beta{+}n)\Gamma(d/2{-}\alpha{+}n)\Gamma({}2\alpha{-}n{-}d/2)}{\Gamma({}\alpha{-}n)\Gamma(d{-}2\alpha{+}n)}\right\}.
\end{eqnarray}
Here, we can use formula (\ref{minn}) of the Appendix \ref{ApA} and finally we obtain the function $f_2(\alpha,\beta, q_1)$ in the form:
%\begin{equation}
%\Gamma(\tau-n)=(-1)^n\frac{\Gamma(\tau)\Gamma(1-\tau)}{\Gamma(1-\tau+n)}
%\end{equation}
\begin{eqnarray}
f_2(\alpha,\beta,q_1)&=&r^{\frac{d}{2}{-}2\alpha{-}\beta}
\frac{S_d\Gamma(\frac{d}{2}{-}2\alpha)\Gamma(\beta{+}2\alpha{-}\frac{d}{2})}{2 \Gamma(\beta) }{_3F_2\left[\left.\begin{array}{c}
\beta{+}2\alpha{-}\frac{d}{2},1{+}\alpha{-}\frac{d}{2},{}\alpha\\1{+}2\alpha{-}\frac{d}{2}, \frac{d}{2}\end{array}\right|{-}\frac{q^2_1}{r}\right]}{+}
\nonumber\\&&r^{{-}\beta}(q^2_1)^{\frac{d}{2}{-}2\alpha}\frac{S_d\Gamma(\frac{d}{2})\Gamma({-}\frac{d}{2}{+}2\alpha)\Gamma^2(\frac{d}{2}{-}\alpha)}{2\Gamma^2({}\alpha)\Gamma(d{-}2\alpha) }{_3F_2\left[\left.\begin{array}{c}
\frac{d}{2}{-}\alpha,\beta,1{-}\alpha\\d{-}2\alpha,1{-}2\alpha{+}\frac{d}{2}\end{array}\right|{-}\frac{q^2_1}{r}\right]},
\end{eqnarray}
where functions ${_3F_2}$ are the generalised hypergeometric functions~\cite{Erdeley}, defined  in (\ref{ghyper}) of Appendix~\ref{ApA}. Note that  the second term disappears at $\alpha=0$ and  the function ${_3F_2}$ in first term is equal to unity. Therefore again the result of usual $O(N)$-symmetric $\phi^4$ theory \cite{rgbooks} is recovered.

Taking in mind to calculate integral (\ref{int4}) we should consider the case  $\beta=2$:
\begin{eqnarray}
f_2(\alpha,2,q_1)&=&r^{d/2{-}2\alpha{-}2}
\frac{S_d\Gamma(\frac{d}{2}{-}2\alpha)\Gamma(2{+}2\alpha{-}\frac{d}{2})}{2  }{_3F_2\left[\left.\begin{array}{c}
2{+}2\alpha{-}\frac{d}{2},1{+}\alpha{-}\frac{d}{2},{}\alpha\\1{+}2\alpha{-}\frac{d}{2}, \frac{d}{2}\end{array}\right|{-}\frac{q^2_1}{r}\right]}{+}
\nonumber\\&&r^{{-}2}(q^2_1)^{\frac{d}{2}{-}2\alpha}\frac{S_d\Gamma(\frac{d}{2})\Gamma({-}\frac{d}{2}{+}2\alpha)\Gamma^2(\frac{d}{2}{-}\alpha)}{2\Gamma^2({}\alpha)\Gamma(d{-}2\alpha) }{_3F_2\left[\left.\begin{array}{c}
\frac{d}{2}{-}\alpha,2,1{-}\alpha\\d{-}2\alpha,1{-}2\alpha{+}\frac{d}{2}\end{array}\right|{-}\frac{q^2_1}{r}\right]}.\label{ff2}
\end{eqnarray}

In the first function ${_3F_2}$  the first numerator parameter is equal to the first denominator parameter plus one (for definition of numerator and denominator parameters see (\ref{ghyper})), therefore using definition for generalised hypergeometric function this function  can be rewritten as the sum of two ${_2F_1}$ functions. As final expression  for $f_2(\alpha, 2, q_2)$ we have:
\begin{eqnarray}\label{f2fin}
f_2(\alpha,2,q_1)&=&r^{d/2{-}2\alpha{-}2}
\frac{S_d\Gamma(\frac{d}{2}{-}2\alpha)\Gamma(2{+}2\alpha{-}\frac{d}{2})}{2  }\left({_2F_1\left[\left.\begin{array}{c}
1{+}\alpha{-}\frac{d}{2},{}\alpha\\ \frac{d}{2}\end{array}\right|{-}\frac{q^2_1}{r}\right]}\right.{-}\nonumber\\&&\left.\frac{q^2_1}{r}\frac{{}\alpha(1{+}\alpha{-}\frac{d}{2})}{(1{+}2\alpha{-}\frac{d}{2})\frac{d}{2}}{_2F_1\left[\left.\begin{array}{c}
2{+}\alpha{-}\frac{d}{2},1{+}\alpha\\ 1{+}\frac{d}{2}\end{array}\right|{-}\frac{q^2_1}{r}\right]}\right){+}
\nonumber\\&&r^{{-}2}(q^2_1)^{\frac{d}{2}{-}2\alpha}\frac{S_d\Gamma(\frac{d}{2})\Gamma({-}\frac{d}{2}{+}2\alpha)\Gamma^2(\frac{d}{2}{-}\alpha)}{2\Gamma^2({}\alpha)\Gamma(d{-}2\alpha) }{_3F_2\left[\left.\begin{array}{c}
\frac{d}{2}{-}\alpha,2,1{-}\alpha\\d{-}2\alpha,1{-}2\alpha{+}\frac{d}{2}\end{array}\right|{-}\frac{q^2_1}{r}\right]}.
\end{eqnarray}
Here, similarly as for function $f_1(\alpha,\beta,q_1)$  a particular case $\alpha=0$ gives the  known result, because the first $_2F_1$ function becomes unity and the two last terms in (\ref{f2fin}) disappear.

Note that function $f_2(\alpha,\beta,q_1)$ is a particular case of integral (25) of Ref.~\cite{BoosDavydychev} with two different external momenta, when one of them is zero. In  Ref.~\cite{BoosDavydychev} result was obtained as a combination of two Lauricella generalized functions  of three variables. When one puts one  external momentum to be zero that result can be reduced to Kamp\'e de F\'eriet functions~\cite{Exton} of two variables. We obtain a simpler expression in a form of a combination of two generalized hypergeometric functions $_3F_2$. Therefore our result (\ref{f2fin})  can be used to find possible reductions for more general hypergeometric functions.

Let us perform now the calculation of function $f_3(\alpha,\beta,\gamma, q_1)$. First with the help of Feynman parametrisation (\ref{fenpam}) we get an integral  with massive propagators only:
\begin{equation}\label{fey_par2}
f_3(\alpha,\beta,\gamma,q_1)=\frac{\Gamma(\beta+\alpha)}{\Gamma(\alpha)\Gamma(\beta)}\int_0^1 dx x^{\beta-1}(1-x)^{\alpha-1}\int_{q_2}\frac{1}{(rx+q^2)^{\beta+\alpha}(r+({\bf q}_1+{\bf q}_2)^2)^\gamma}.
\end{equation}
Then we can use expression (20) of Ref. \cite{BoosDavydychev} for the integral with two propagators having different masses. In our notation this formula has following form:
\begin{eqnarray}\label{boos}
\int_{q}\!\!\frac{1}{(m^2_1{+}q^2)^a(m^2_2{+}({\bf k}{+}{\bf q})^2)^b}&{=}&\frac{S_d\Gamma(\frac{d}{2})\Gamma(a{-}\frac{d}{2})}{2\Gamma(a)}(m^2_1)^{d/2{-}a{-}b}
 \!\left\{\!\frac{\Gamma(a)\Gamma(d/2{-}a)\Gamma(a{+}b{-}d/2)}{\Gamma(a{-}d/2)\Gamma(d/2)\Gamma(b)}\right.{\times} \nonumber\\&& F_4\left[a,a{+}b{-}d/2;d/2,a{-}d/2{+}1|{-}\frac{k^2}{m_2^2},\frac{m_1^2}{m_2^2}\right]{+}
\nonumber\\&&
\left.\left(\frac{m_1^2}{m_2^2}\right)^{d/2{-}a}F_4\left[b,d/2;d/2,d/2{-}a{+}1|{-}\frac{k^2}{m_2^2},\frac{m_1^2}{m_2^2}\right]\right\},
\end{eqnarray}
where functions $F_4$  are hypergeometric Appell functions of two variables~\cite{Erdeley}. Definition of Appell function $F_4$ is given in (\ref{appell}) the Appendix~\ref{ApA}.
Comparing this formula with the last integral in (\ref{fey_par2})  one can see that in our case $m_1^2=rx$, $m_2^2=r$, $a=1+\alpha$ and $b=1$. Substituting (\ref{boos}) into (\ref{fey_par2}) we get:
\begin{eqnarray}\label{f3afterq}
f_3(\alpha,\beta,\gamma,q_1)&{=}&\frac{S_d\Gamma(d/2)\Gamma(\beta{+}\alpha{-}d/2)}{2\Gamma({}\alpha)\Gamma(\beta)}\,r^{d/2{-}\alpha{-}\beta{-}\gamma}
\left\{\frac{\Gamma(\beta{+}\alpha)\Gamma(d/2{-}\alpha{-}\beta)\Gamma(\gamma{+}\beta{+}\alpha{-}d/2)}{\Gamma(\beta{+}\alpha{-}d/2)\Gamma(d/2)\Gamma(\gamma)}{\times}\right. \nonumber\\&& \int_0^1 dx x^{\beta-1}(1-x)^{\alpha-1}F_4\left[\beta{+}\alpha,\beta{+}\gamma{+}\alpha{-}d/2;d/2,1{+}\beta{+}\alpha{-}d/2|{-}\frac{q_1^2}{r},x\right]{+}
\nonumber\\&&
\left.\int_0^1 dx (1-x)^{{}\alpha{-}1} x^{d/2{-}\alpha{-}1}F_4\left[\gamma,d/2;d/2,d/2{-}\alpha{-}\beta+1|{-}\frac{q_1^2}{r},x\right]\right\}.
\end{eqnarray}
Therefore we need to perform now only the integration over the Feynman parameter $x$. Using the definition of the Appell function $F_4$ we can rewrite it as an infinite sum of functions $_2F_1$:
\begin{equation}
F_4[a,b;c,c';x,y]=\sum_{n>0}\frac{x^n}{n!}\frac{(a)_n(b)_n}{(c)_n}{_2F_1}\left[\left.\begin{array}{c}
a+n,b+n\\ c'\end{array}\right|y\right].
\end{equation}
Then substituting this expression into (\ref{f3afterq}) we are able to use table integral for $_2F_1$ functions. As a consequence, using formula 7.512.5 from \cite{GradshteynRyzhik} %for first term and formula 7.512.4  \cite{GradshteynRyzhik} for second term
we obtain result in the form:
\begin{eqnarray}\label{f3fin}
f_3(\alpha,\beta,\gamma,q_1)&{=}&\frac{S_d\Gamma(d/2)}{2}r^{d/2{-}\alpha{-}\beta-\gamma}
\sum_{n>0}\left({-}\frac{q_1^2}{r}\right)^n\frac{1}{n!}\left\{\frac{\Gamma(d/2-\alpha{-}\beta)\Gamma(\beta+\gamma+\alpha{-}d/2)}{\Gamma(d/2)\Gamma(\gamma)}\times\right. \nonumber\\&&  \frac{(\beta{+}\alpha)_n(\beta{+}\gamma{+}\alpha{-}d/2)_n}{(d/2)_n}
{_3F_2}\!\left[\left.\begin{array}{c}
\beta{+}\alpha{+}n,\beta{+}\gamma{+}\alpha{-}d/2{+}n,\beta\\1{+}\beta{+}\alpha{-}d/2,\beta{+}\alpha\end{array}\right|1\right]{+}
\nonumber\\&&
\left.\frac{\Gamma(\beta{+}\alpha{-}d/2)\Gamma(d/2{-}\alpha)}{\Gamma(d/2)}
(\gamma)_n \,\,{_3F_2}\!\left[\left.\begin{array}{c}
\gamma{+}n,d/2{+}n,d/2{-}\alpha\\d/2{-}\alpha{-}\beta{+}1,d/2\end{array}\right|1\right]\right\}.
\end{eqnarray}
We want to note here that the  function $f_3(\alpha,\beta,\gamma,q_1)$ is a particular case of integral (35) of Ref.~\cite{BoosDavydychev} with two different external momenta, when one of them is zero. Result of evaluation of such integral with both non-zero external  momenta is the complex Lauricella generalized function ~\cite{BoosDavydychev}, which can be reduced to Kamp\'e de F\'eriet function when one of those momenta is put to be zero. Both arguments of obtained Kamp\'e de F\'eriet function are not unit. Here we obtain result as two infinite sums of $_3F_2$  functions with unit argument, which makes possible to use known relations for  $_3F_2$  functions with unit argument.

 Let us consider the case $\beta=\gamma=1$, which we need for integrals (\ref{int3}), (\ref{int6}), (\ref{int7}). For these values of parameters we have:
\begin{eqnarray}\label{f3par1}
f_3(\alpha,1,1,q_1)&=&\frac{S_d\Gamma(d/2)}{2}r^{d/2{-}\alpha{-}2}
\left\{\frac{\Gamma(d/2-\alpha{-}1)\Gamma(2+\alpha{-}d/2)}{\Gamma(d/2)}\times\right. \nonumber\\&&  \sum_{n>0}\left(-\frac{q_2^2}{r}\right)^n\frac{1}{n!}\frac{(1{+}\alpha)_n(2{+}\alpha{-}d/2)_n}{(d/2)_n}
{_3F_2}\!\left[\left.\begin{array}{c}
1{+}\alpha{+}n,2{+}\alpha{-}d/2{+}n,1\\2{+}\alpha{-}d/2,1{+}\alpha\end{array}\right|1\right]{+}
\nonumber\\&&
\left.{\Gamma(1{+}\alpha{-}d/2)\Gamma(d/2{-}\alpha)}\sum_{n>0}\left(-\frac{q_2^2}{r}\right)^n\frac{\Gamma(1{+}n)}{n!}
\frac{\Gamma({-}1{-}2n)}{\Gamma(d/2{-}1{-}n)\Gamma({-}n)}\right\}.
\end{eqnarray}
Consider first the $_3F_2$ function in the second line of (\ref{f3par1}).
Using Thomae transformation for $_3F_2$ functions with unit argument (see e.g. formula 4.3.1 in \cite{Slater}) we  obtain:
\begin{eqnarray}
{_3F_2}\left[\left.\begin{array}{c}
1{+}\alpha{+}n,2{+}\alpha{-}d/2{+}n,1\\2{+}\alpha{-}d/2,1{-}\alpha\end{array}\right|1\right]&{=}&
\frac{\Gamma(2{+}\alpha{-}d/2)\Gamma(1{+}\alpha)\Gamma({-}1{-}2n)}{\Gamma({}\alpha{-}n)\Gamma(1{-}d/2{+}\alpha{-}n)}\times\nonumber\\&&\label{3f2}
{_3F_2}\left[\left.\begin{array}{c}
{}\alpha,1{+}\alpha{-}d/2,{-}1{-}2n\\{}\alpha{-}n,1{+}\alpha{-}d/2{-}n\end{array}\right|1\right].
\end{eqnarray}
Note, that in the right hand side of Eq. (\ref{3f2}) the sum of the denominator parameters of $_3F_2$ function exceeds sum of its numerator parameters by one. It makes possible to use the  Saalschutz's theorem for its transformation   (see e.g. formula  2.3.1.4 in \cite{Slater}):
 \begin{equation}\label{3f2d}
{_3F_2}\left[\left.\begin{array}{c}
{}\alpha,1{+}\alpha{-}d/2,{-}1{-}2n\\{}\alpha{-}n,1{+}\alpha{-}d/2{-}n\end{array}\right|1\right]{=}
\frac{\Gamma(\alpha-n)\Gamma(d/2{+}n)\Gamma(1{+}n)\Gamma(-\alpha{+}d/2-1-n)}{\Gamma(\alpha{-}d/2{+}n)\Gamma(-n)\Gamma(d/2-1-n)\Gamma(\alpha{+}1{+}n)}.
\end{equation}
Combining (\ref{f3par1}) with (\ref{3f2}) and (\ref{3f2d}) and using for $\Gamma(-1-2n)$ the duplication formula (see  (\ref{dupl}) in Appendix~\ref{ApA}) we get result for $f_3(q_2,1,1,\alpha)$ in the following form:
\begin{eqnarray}\label{f3fin1}
f_3(\alpha,1,1,q_1)&=&\frac{S_d\Gamma(\frac{d}{2})}{2}r^{\frac{d}{2}{-}\alpha{-}2}
\left\{-\frac{\Gamma(\frac{d}{2}-\alpha{-}1)^2\Gamma(2+\alpha{-}\frac{d}{2})^2}{2\Gamma(\frac{d}{2}-1)\Gamma(1-\frac{d}{2}+\alpha)\Gamma(\frac{d}{2}-\alpha)}
{_2F_1}\left[\left.\begin{array}{c}
1,2-\frac{d}{2}\\3/2\end{array}\right|-\frac{q_1^2}{4r}\right]\right.
\nonumber\\&&
\left.-\frac{\Gamma(1+\alpha-\frac{d}{2})\Gamma(\frac{d}{2}-\alpha)}{2 \Gamma(\frac{d}{2}-1)}{_2F_1}\left[\left.\begin{array}{c}
1,2-\frac{d}{2}\\3/2\end{array}\right|-\frac{q_1^2}{4r}\right]\right\}=\nonumber\\&&
\frac{S_d(\frac{d}{2}-1)}{2}\Gamma(\frac{d}{2}-\alpha-1)\Gamma(2+\alpha-\frac{d}{2})r^{\frac{d}{2}{-}\alpha{-}2}{_2F_1}\left[\left.\begin{array}{c}
1,2-\frac{d}{2}\\3/2\end{array}\right|-\frac{q_1^2}{4r}\right].
\end{eqnarray}

At the evaluation of $f_4(\alpha,\beta,\gamma)$ we again can use result (\ref{f1}). To do that we apply Feynman parametrisation as well as use Mellin-Barnes representation for massive propagator, presenting   $f_4(\alpha,\beta,\gamma)$  in the form:
\begin{eqnarray}
f_4(\alpha,\beta,\gamma,q_1)&{=}&\frac{1}{(2\pi i)}\frac{\Gamma(\gamma{+}\alpha)}{\Gamma(\gamma)\Gamma({}\alpha)}\int_{-i\infty}^{i\infty} ds (r)^{s}\Gamma({-}s)\Gamma(\beta{+}s){\times}\nonumber\\&&\int_0^1 dx \int_{q_2}\frac{x^{\gamma{-}1}(1{-}x)^{{}\alpha{-}1}}{(q_2^2)^{\beta{+}\alpha{+}s}(rx{+}({\bf q}_1{+}{\bf q}_2)^2)^{\gamma{+}\alpha}}.
\end{eqnarray}
Integration over $q_1$ gives the $_2F_1$ function. To make the integration over Feynman parameter $x$ we rewrite it in the Mellin-Barnes representation. Then the integration over $x$ results in:
\begin{eqnarray}
f_4(\alpha,\beta,\gamma,q_1)&{=}&r^{d/2{-}2\alpha{-}\beta{-}\gamma}\frac{S_d \Gamma(d/2)}{2\Gamma(\gamma)}
\frac{1}{(2\pi i)^2}\int_{{-}i\infty}^{i\infty}\!\! ds\int_{{-}i\infty}^{i\infty} \!\!dt\Gamma({-}s)\Gamma(\beta{+}s)\Gamma({-}t)\Gamma(\beta{+}\alpha{+}s{+}t){\times}\nonumber\\&&
\frac{\Gamma(d/2{-}\alpha{-}\beta{-}s)\Gamma(\beta+\gamma{{+}}2\alpha{{-}}d/2{+}s{+}t)\Gamma({-}\beta-2\alpha+d/2{-}s{-}t)}{\Gamma({{}}\alpha{+}\beta{+}s)
\Gamma(d/2{-}\beta-\alpha{-}s{-}t)
\Gamma(d/2{+}t)}\left(\frac{q^2_1}{r}\right)^{t}\!\!.
\end{eqnarray}
Closing the contour of integration to the right  we use the set of the residues of the functions  $\Gamma(-t)$ and $\Gamma(d/2-2\alpha-\beta-s-t)$ and obtain the result in the form:
\begin{eqnarray}\label{f4fp}
f_4(\alpha,\beta,\gamma,q_1)&{=}&r^{d/2{-}2\alpha{-}\beta{-}\gamma}\frac{S_d \Gamma(d/2)}{2}
\frac{1}{(2\pi i)}{\times}\nonumber\\&&\!\!\!\int_{-i\infty}^{i\infty}\!\!ds\sum_{n\ge 0}\frac{1}{n!}\left\{\!\left({-}\frac{q^2_1}{r}\right)^{n}\!\Gamma({-}s)\Gamma(\beta{+}s)\Gamma(\beta{+}\alpha{+}s{+}n)\Gamma(d/2{-}2\alpha{-}\beta{-}s)\,
{\times}\right.\nonumber\\&&
\frac{
\Gamma(1{+}\beta{-}d/2{+}2\alpha{+}s)\Gamma(1{+}\beta{-}d/2{+}\alpha{+}s{+}n)
\Gamma(\beta{+}\gamma{+}2\alpha{-}d/2{+}s{+}n)}{
\Gamma({}\alpha{+}\beta{+}s)\Gamma(d/2{+}n)\Gamma(1{+}\beta{-}d/2{+}\alpha{+}s)\Gamma(1{+}\beta{+}2\alpha{-}d/2{+}s{+}n)}\,{+}
\nonumber\\&&
\left(\frac{q^2_1}{r}\right)^{d/2-2\alpha-\beta-s+n}(-1)^n
\frac{\Gamma({-}s)\Gamma(\beta{+}s)\Gamma(d/2{-}\alpha{-}\beta{-}s)\Gamma(\gamma{+}n)}{\Gamma(1{-}\beta{-}2\alpha{+}d/2{-}s{+}n)}\times\nonumber\\&&
\left.\frac{
\Gamma(d/2{-}\alpha{+}n)\Gamma(\beta{+}2\alpha{-}d/2{+}s)
\Gamma(1{-}\beta{-}2\alpha{+}d/2{-}s)\Gamma(1{-}\alpha{+}n)}{
\Gamma({}\alpha{+}\beta{+}s)\Gamma({}\alpha)\Gamma(1{-}\alpha)\Gamma(d{-}2\alpha{-}\beta{-}s{+}n)}
\right\}.
\end{eqnarray}
Note that  formula (\ref{minn}) was used to obtain (\ref{f4fp}).
Performing the next integration we close the contour to the right for the  first term (using poles of  $\Gamma(-s)$ and $\Gamma(d/2-2\alpha-\beta-s)$), while for  the second term we close the contour to the left (using poles $\Gamma(\beta+s)$ and $\Gamma(-d/2+2\alpha+\beta+s)$). Then using formula (\ref{minn}) our result can be presented via Kamp\'e de F\'eriet functions (see (\ref{kampe}) in Appendix \ref{ApA}):
\begin{eqnarray}
f_4({\alpha},{\beta},{\gamma},q_1)&{=}&\frac{r^{\frac{d}{2}{-}2\alpha{-}2} S_d \Gamma(\frac{d}{2})}{2}
\left\{\frac{\Gamma(\frac{d}{2}{-}2\alpha{-}\beta)\Gamma(\beta)\Gamma(\beta{+}\gamma{+}2\alpha{-}\frac{d}{2})}{\Gamma(d/2)}{\times}\right.\nonumber\\&&
{F_{1,2,1}^{3,1,0}\!\left[\left.\begin{array}{c}
1{+}\beta{+}\alpha{-}\frac{d}{2},\gamma{+}\beta{+}2\alpha{-}\frac{d}{2},\beta{+}\alpha;\beta; 0\\
1{+}\beta{+}2\alpha{-}\frac{d}{2};\beta{+}\alpha,1{+}\beta{+}\alpha{-}\frac{d}{2};\frac{d}{2}\end{array}\right|1,{-}\frac{q^2_1}{r}\right]}
+\nonumber\\&&
\frac{\Gamma(\beta-\frac{d}{2}{+}2\alpha)\Gamma(\gamma)\Gamma(\frac{d}{2}{-}2\alpha)}{\Gamma(d/2)}\nonumber\\&&
{F_{1,3,1}^{3,2,0}\left[\!\!\left.\begin{array}{c}
\gamma,-\alpha+\frac{d}{2},1-\alpha;1,\frac{d}{2}-2\alpha; 0\\
1;1-\alpha,-\alpha+\frac{d}{2},1-\beta+\frac{d}{2}-2\alpha;\frac{d}{2}\end{array}\right|1,-\frac{q^2_1}{r}\right]}
+
\nonumber\\&&
\left(\frac{q^2_1}{r}\right)^{\frac{d}{2}{-}2\alpha}\!\!
\frac{\Gamma(\beta)\Gamma(\gamma)\Gamma^2(\frac{d}{2}{-}\alpha)\Gamma({-}\frac{d}{2}{+}2\alpha)}{\Gamma^2(\alpha)\Gamma(d{-}2\alpha)}\times\nonumber\\
&&\nonumber
{F_{2,0,0}^{0,3,3}\left[\left.\begin{array}{c}
0;\beta,\frac{d}{2}{-}\alpha,1{-}\alpha; \gamma,\frac{d}{2}{-}\alpha,1{-}\alpha\\
1-2\alpha+\frac{d}{2},d-2\alpha;0;0\end{array}\right|{-}\frac{q^2_1}{r},{-}\frac{q^2_1}{r}\right]}{+}
\nonumber\\&&
\frac{\Gamma(\frac{d}{2}{-}2\alpha)\Gamma(\gamma)\Gamma(\beta{+}2\alpha{-}\frac{d}{2})}{\Gamma(d/2)}\times\nonumber\\&&\left.
{F_{2,1,0}^{0,4,3}\left[\left.\begin{array}{c}
0;1,{}\alpha,1{+}\alpha{-}\frac{d}{2},\beta{+}2\alpha{-}\frac{d}{2}; \gamma,\frac{d}{2}{-}\alpha,1{-}\alpha\\
1,d/2;1{+}2\alpha{-}\frac{d}{2};0\end{array}\right|{-}\frac{q^2_1}{r},{-}\frac{q^2_1}{r}\right]}
\right\}. \label{gf4}
\end{eqnarray}
To calculate integral (\ref{int9}), we have to consider  $\beta=\gamma=1$, in this case expression (\ref{gf4})  is reduced to the form:
\begin{eqnarray}
f_4({\alpha},{1},{1},q_1)&{=}&\frac{r^{d/2{-}2\alpha{-}2} S_d \Gamma(d/2)}{2}
\left\{\frac{\Gamma(\frac{d}{2}{-}2\alpha{-}1)\Gamma(2{+}2\alpha{-}\frac{d}{2})}{\Gamma(d/2)}\right.
\times\nonumber\\&&
{F_{0,2,1}^{2,1,0}\!\left[\left.\begin{array}{c}
2{+}\alpha{-}\frac{d}{2},1{+}\alpha;1; 0\\
0;1{+}\alpha,2{+}\alpha{-}\frac{d}{2};\frac{d}{2}\end{array}\right|1,{-}\frac{q^2_1}{r}\right]}
+\nonumber\\&&
\frac{\Gamma(1-d/2{+}2\alpha)\Gamma(d/2{-}2\alpha)}{\Gamma(d/2)}
{F_{0,2,1}^{2,1,0}\left[\left.\begin{array}{c}
-\alpha+d/2,1-\alpha;1; 0\\
0;1-\alpha,-\alpha+d/2;d/2\end{array}\right|1,-\frac{q^2_1}{r}\right]}
+
\nonumber\\&&
\left(\frac{q^2_1}{r}\right)^{\frac{d}{2}{-}2\alpha}\frac{\Gamma^2(\frac{d}{2}{-}\alpha)\Gamma({-}\frac{d}{2}{+}2\alpha)}{\Gamma^2(\alpha)\Gamma(d{-}2\alpha)}
\times\nonumber\\&&{F_{2,0,0}^{0,3,3}\left[\left.\begin{array}{c}
0;1,\frac{d}{2}{-}\alpha,1{-}\alpha; 1,\frac{d}{2}{-}\alpha,1{-}\alpha\\
1-2\alpha+\frac{d}{2},d-2\alpha;0;0\end{array}\right|{-}\frac{q^2_1}{r},{-}\frac{q^2_1}{r}\right]}{+}
\nonumber\\&&
\frac{\Gamma(d/2{-}2\alpha)\Gamma(1{+}2\alpha{-}d/2)}{\Gamma(d/2)}\times\nonumber\\&&
\left.{F_{2,0,0}^{0,3,3}\left[\left.\begin{array}{c}
0;1,{}\alpha,1{+}\alpha{-}d/2; 1,d/2{-}\alpha,1{-}\alpha\\
1,d/2;0;0\end{array}\!\!\right|{-}\frac{q^2_1}{r},{-}\frac{q^2_1}{r}\right]}
\right\}.
\end{eqnarray}
Here further reduction can be performed. The first two terms can be presented as an infinite sum of $_3F_2$ functions with unit argument. Using for these functions Thomae transformation  (see e.g. formula 4.3.1 in \cite{Slater}) and  Saalschutz's theorem (see e.g. formula  2.3.1.4 in \cite{Slater}) as well as formulae (\ref{dupl}) and (\ref{minn}) from Appendix \ref{ApA} finally we get the following result:
\begin{eqnarray}
f_4({\alpha},{1},{1},q_1)&{=}&\frac{r^{\frac{d}{2}{-}2\alpha{-}2} S_d \Gamma(\frac{d}{2})}{2}\!
\left\{\frac{(\frac{d}{2}{-}1)\Gamma(\frac{d}{2}{-}2\alpha{-}1)\Gamma(2{+}2\alpha{-}\frac{d}{2})}{ \Gamma(\frac{d}{2})}{_2F_1}\!\left[\!\!\left.\begin{array}{c}
1,2{-}\frac{d}{2}\\3/2\end{array}\right|{-}\frac{q_1^2}{4r}\right]\right.
{+}
\nonumber\\&&
\!\!\!\!\!\!\!\!\left(\frac{q^2_1}{r}\right)^{\frac{d}{2}{-}2\alpha}\frac{\Gamma^2(\frac{d}{2}{-}\alpha)\Gamma({-}\frac{d}{2}{+}2\alpha)}{\Gamma^2(\alpha)
\Gamma(d{-}2\alpha)}
{F_{2,0,0}^{0,3,3}\left[\!\!\!\left.\begin{array}{c}
0;1,\frac{d}{2}{-}\alpha,1{-}\alpha; 1,\frac{d}{2}{-}\alpha,1{-}\alpha\\
1-2\alpha+\frac{d}{2},d-2\alpha;0;0\end{array}\right|{-}\frac{q^2_1}{r},{-}\frac{q^2_1}{r}\!\!\right]}{+}
\nonumber\\&&
\!\!\!\!\!\!\left.
\frac{\Gamma(\frac{d}{2}{-}2\alpha)\Gamma(1{+}2\alpha{-}\frac{d}{2})}{\Gamma(d/2)}
{F_{2,0,0}^{0,3,3}\left[\left.\begin{array}{c}
0;1,{}\alpha,1{+}\alpha{-}\frac{d}{2}; 1,\frac{d}{2}{-}\alpha,1{-}\alpha\\
1,\frac{d}{2};0;0\end{array}\!\!\right|{-}\frac{q^2_1}{r},{-}\frac{q^2_1}{r}\right]}
\right\}.\label{ff4}
\end{eqnarray}

Expressions (\ref{f1}), (\ref{ff2}), (\ref{f3fin1}), (\ref{ff4}) obtained in this section
for functions (\ref{fan1})-(\ref{fan4}) will be further used for the calculation of  the two-loop integrals. It is done in the following section.

\section{Expressions for two-loop integrals}\label{IV}
After performing   an integration over $q_2$ in integrals(\ref{int1})-(\ref{int2}), expressions for them have only one
%To find two-loop massive  integrals presented in Section~\ref{II} we should perform
momentum integration over $q_1$.  Substituting  functions (\ref{fan1})-(\ref{fan4}) into (\ref{int1})-(\ref{int2}) these integrals can be written in the following form:
\begin{equation}
I(1,1,0,2,0,2)=\int_{q_1}d q_1\frac{(q_1^2)^{\frac{a-d}{2}}}{(r+q^2_1)^2}f_1\left(\frac{d-a}{2},2,q_1^2\right),
\end{equation}
\begin{equation}
I(1,1,0,3,0,1)=\int_{q_1}d q_1\frac{(q_1^2)^{\frac{a-d}{2}}}{(r+q^2_1)^3}f_1\left(\frac{d-a}{2},1,q_1^2\right),
\end{equation}
\begin{equation}
I(1,1,0,2,1,1)=\int_{q_1}d q_1\frac{(q_1^2)^{\frac{a-d}{2}}}{(r+q^2_1)^2}f_3\left(\frac{d-a}{2},1,1,q_1^2\right),
\end{equation}
\begin{equation}
I(1,1,1,2,0,2)=\int_{q_1}d q_1\frac{(q_1^2)^{\frac{a-d}{2}}}{(r+q^2_1)^2}f_2\left(\frac{d-a}{2},2,q_1^2\right),
\end{equation}
\begin{equation}
I(2,1,0,3,0,1)=\int_{q_1}d q_1\frac{(q_1^2)^{a-d}}{(r+q^2_1)^3}f_1\left(\frac{d-a}{2},1,q_1^2\right),
\end{equation}
\begin{equation}
I(2,1,0,2,1,1)=\int_{q_1}d q_1\frac{(q_1^2)^{a-d}}{(r+q^2_1)^2}f_3\left(\frac{d-a}{2},1,1,q_1^2\right),
\end{equation}
\begin{equation}
I(1,2,0,2,1,1)=\int_{q_1}d q_1\frac{(q_1^2)^{\frac{a-d}{2}}}{(r+q^2_1)^2}f_3\left({d-a},1,1,q_1^2\right),
\end{equation}
\begin{equation}\label{int110112}
I(1,1,0,1,1,2)=\int_{q_1}d q_1\frac{(q_1^2)^{\frac{a-d}{2}}}{(r+q^2_1)}f_3\left(\frac{d-a}{2},1,2,q_1^2\right),
\end{equation}
\begin{equation}\label{int111211}
I(1,1,1,2,1,1)=\int_{q_1}d q_1\frac{(q_1^2)^{\frac{a-d}{2}}}{(r+q^2_1)^2}f_4\left(\frac{d-a}{2},1,1,q_1^2\right).
\end{equation}
Integrands in the integrals presented above are functions of  absolute value of $q_1$ therefore integration over angular part  can be performed separately. For the integrals including $_2F_1$  (that is with functions (\ref{fan1}) and (\ref{fan2}) ) the table integrals can be used (see e. g. 2.21.1.15 in \cite{PrudnikovBrychkovMarichev}). Therefore  we  get the following expressions for (\ref{int1})-(\ref{int3}), (\ref{int5})-(\ref{int7}):

\begin{eqnarray}
I(1,1,0,2,0,2)&{=}&r^{a-4}\left(\frac{S_d}{2}\right)^2\Bigg\{\frac{\Gamma(\frac{a}{2})\Gamma(\frac{d}{2})\Gamma^2(2-\frac{a}{2})\Gamma(a-2)}{\Gamma(\frac{d+a}{2}-2)}\left(1+\frac{(2-\frac{a}{2})(\frac{a-d}{2})}{(\frac{a}{2}-1)(3-a)}\right)+\nonumber\\
\!\!&&\frac{\Gamma(\frac{a}{2})\Gamma(\frac{d}{2})\Gamma(2-\frac{2a-d}{2})\Gamma(4-a)\Gamma(\frac{a}{2}-2)}{\Gamma(\frac{d-a}{2})\Gamma(2-\frac{a-d}{2})}\,{_3F_2\!\!\left[\!\left.\begin{array}{c}
2-\frac{2a-d}{2},4-a,2\\
2-\frac{a-d}{2},3-\frac{a}{2}\end{array}\right|1\right]}\!\Bigg\},
\end{eqnarray}

\begin{eqnarray}
I(1,1,0,3,0,1)&{=}&r^{a{-}4}\left(\frac{S_d}{2}\right)^2\Bigg\{\frac{\Gamma(\frac{a}{2})\Gamma(\frac{d}{2})\Gamma(1{-}\frac{a}{2})\Gamma(3{-}\frac{a}{2})
\Gamma(a{-}1)}{2\Gamma(\frac{d{+}a}{2}{-}1)}\times\nonumber\\&&
\left(1\!{+}
\frac{{}\frac{d{-}a}{2}}{(a{-}2)}\!\left(2\frac{(1{-}\frac{a}{2})}{(\frac{a}{2}{-}2)}{+}\frac{(1{-}\frac{a{-}d}{2})}{(a{-}3)}\!\right)\!\right)+\nonumber\\
&&\frac{\Gamma(\frac{a}{2})\Gamma(\frac{d}{2})\Gamma(3{-}\frac{2a-d}{2})\Gamma(4{-}a)\Gamma(\frac{a}{2}{-}3)}{\Gamma(\frac{d-a}{2})\Gamma(3-\frac{a-d}{2})}{_3F_2\left[\left.\begin{array}{c}
3{-}\frac{2a-d}{2},4{-}a,3\\
3{-}\frac{a-d}{2},4{-}\frac{a}{2}\end{array}\right|1\right]}\Bigg\},
\end{eqnarray}

\begin{eqnarray}\label{i110211}
I(1,1,0,2,1,1)&{=}&r^{a{-}4}\left(\frac{S_d}{2}\right)^2\Bigg\{{\left(\frac{d}{2}-1\right)\Gamma(\frac{a}{2})\Gamma(\frac{a}{2}-1)\Gamma^2(2{-}\frac{a}{2})
}\!\Bigg({_2F_1}\left[1,2-\frac{d}{2},\frac{3}{2};\frac{1}{4}\right]{+}\nonumber\\&&
\frac{(2-\frac{d}{2})}{6(\frac{a}{2}-1)}{_2F_1}\left[2,3-\frac{d}{2},\frac{5}{2};\frac{1}{4}\right]\Bigg)+\nonumber\\
&&2^{a-4}\frac{\left(\frac{d}{2}{-}1\right)\Gamma(\frac{a}{2}{-}1)\Gamma(2{-}\frac{a}{2})\Gamma(\frac{3}{2})\Gamma(3{-}\frac{a}{2})\Gamma(4{-}\frac{a+d}{2})
\Gamma(\frac{a}{2}-2)}{\Gamma(2-\frac{d}{2})\Gamma(\frac{7-a}{2})}\times\nonumber\\&&{_2F_1\left[2,4-\frac{a+d}{2},\frac{7-a}{2};\frac{1}{4}\right]}\Bigg\},
\end{eqnarray}

\begin{eqnarray}
I(2,1,0,3,0,1)&{=}&r^{\frac{3a-d}{2}{-}4}\left(\frac{S_d}{2}\right)^2\Bigg\{\frac{\Gamma(\frac{d}{2})\Gamma(1{-}\frac{a}{2})\Gamma(\frac{2a-d}{2})
\Gamma(3-\frac{2a-d}{2})
\Gamma(a{-}1)}{2\Gamma(\frac{d{+}a}{2}{-}1)}\!
{\times}\nonumber\\&&\Bigg(1\!{+}
\frac{{}\frac{d{-}a}{2}(1{-}\frac{a}{2})}{(a{-}2)(\frac{2a-d}{2}{-}2)}\left(2{}{+}\frac{(1{-}\frac{a{-}d}{2})(2-\frac{a}{2})}{(a{-}3)(\frac{2a-d}{2}-1)}\!\right)\!\Bigg)+\nonumber\\
&&{+}\frac{\Gamma(\frac{a}{2})\Gamma(\frac{d}{2})\Gamma(3-\frac{3a-2d}{2})\Gamma(4-\frac{3a-d}{2})\Gamma(\frac{2a-d}{2}-3)}{\Gamma(\frac{d-a}{2})\Gamma(3-({a-d}))}
\times\nonumber\\&&{_3F_2\left[\left.\begin{array}{c}
3-\frac{3a-2d}{2},4-\frac{3a-d}{2},3\\
3-({a-d}),4-\frac{2a-d}{2}\end{array}\right|1\right]}\Bigg\},
\end{eqnarray}

\begin{eqnarray}
I(2,1,0,2,1,1)&{=}&r^{\frac{3a-d}{2}{-}4}\left(\frac{S_d}{2}\right)^2
}\!\Bigg\{{\left(\frac{d}{2}{-}1\right)\Gamma(\frac{a}{2}{-}1)\Gamma(2{-}\frac{a}{2})
\Gamma(\frac{2a{-}d}{2})\Gamma(2{-}\frac{2a{-}d}{2}){\times}\nonumber\\&&
\Bigg({_2F_1}\left[1,2-\frac{d}{2},\frac{3}{2};\frac{1}{4}\right]{+}\frac{(2-\frac{d}{2})}{6(\frac{2a-d}{2}-1)}{_2F_1}\left[2,3-\frac{d}{2},\frac{5}{2};\frac{1}{4}\right]\Bigg)+\nonumber\\
&&2^{2a{-}d{-}4}\frac{\left(\frac{d}{2}{-}1\right)\Gamma(\frac{a}{2}{-}1)\Gamma(2{-}\frac{a}{2})\Gamma(\frac{3}{2})\Gamma(3{-}\frac{2a-d}{2})\Gamma(4{-}{a})
\Gamma(\frac{2a{-}d}{2}{-}2)}{\Gamma(2{-}\frac{d}{2})\Gamma(\frac{7{-}(2a{-}d)}{2})}\times\nonumber\\&&
{_2F_1\left[2,4{-}{a},\frac{7{-}(2a{-}d)}{2};\frac{1}{4}\right]}\Bigg\},
\end{eqnarray}

\begin{eqnarray}
I(1,2,0,2,1,1)&{=}&r^{\frac{3a-d}{2}{-}4}\left(\frac{S_d}{2}\right)^2\Bigg\{{\left(\frac{d}{2}-1\right)\Gamma(\frac{a}{2})\Gamma(2{-}\frac{a}{2})
\Gamma(\frac{2a-d}{2}-1)\Gamma(2-\frac{2a-d}{2})
}\!\nonumber\\&&
\Bigg({_2F_1}\left[1,2-\frac{d}{2},\frac{3}{2};\frac{1}{4}\right]{+}\frac{(2-\frac{d}{2})}{6(\frac{a}{2}-1)}{_2F_1}\left[2,3-\frac{d}{2},\frac{5}{2};\frac{1}{4}\right]\Bigg)+\nonumber\\
&&2^{2a{-}d{-}4}\frac{\left(\frac{d}{2}{-}1\right)\Gamma(\frac{2a-d}{2}{-}1)\Gamma(2{-}\frac{2a-d}{2})\Gamma(\frac{3}{2})\Gamma(3{-}\frac{a}{2})\Gamma(4{-}\frac{a+d}{2})
\Gamma(\frac{a}{2}{-}2)}{\Gamma(2{-}\frac{d}{2})\Gamma(\frac{7{-}a}{2})}\times\nonumber\\&&{_2F_1\left[2,4{-}\frac{a+d}{2},\frac{7{-}a}{2};\frac{1}{4}\right]}\Bigg\}.
\end{eqnarray}

To get the integral  $I(1,1,0,1,1,2)$ (\ref{int8}) instead using formula (\ref{int110112})  we can obtain it in other way. It is obvious that
\begin{eqnarray}\label{form1}
I(1,1,0,1,1,2)&{=}&-\frac{\partial }{\partial r}I(1,1,0,1,1,1)-2I(1,1,0,2,1,1).
\end{eqnarray}
The last term in r.h.s. of (\ref{form1}) is already calculated (\ref{i110211}), while the integral in the first term is expressed as  $I(1,1,0,1,1,1)=\int_{q_1}d q_1\frac{(q_1^2)^{\frac{a-d}{2}}}{(r+q^2_1)}f_3\left(\frac{d-a}{2},1,1,q_1^2\right)$. Performing necessary calculation we get:
\begin{eqnarray}
I(1,1,0,1,1,2)&{=}&r^{a{-}4}(3-a)\left(\frac{S_d}{2}\right)^2\Bigg\{{\left(1-\frac{d}{2}\right)\Gamma^2(\frac{a}{2})
\Gamma^2(1{-}\frac{a}{2})
}{_2F_1}\left[1,2{-}\frac{d}{2},\frac{3}{2};\frac{1}{4}\right]\,{+}\nonumber\\
&&2^{a-2}\frac{\left(\frac{d}{2}{-}1\right)\Gamma^2(\frac{a}{2}{-}1)\Gamma^2(2{-}\frac{a}{2})\Gamma(\frac{3}{2})\Gamma(3{-}\frac{a+d}{2})
}{\Gamma(2{-}\frac{d}{2})\Gamma(\frac{5{-}a}{2})}{_2F_1\left[1,3{-}\frac{a+d}{2},\frac{5{-}a}{2};\frac{1}{4}\right]}\Bigg\}{-}\nonumber\\&&
2\times\mbox{r.h.s.}(\protect\ref{i110211})
\end{eqnarray}

To evaluate integral $I(1,1,1,2,0,2)$, {\em i. e.}  to integrate  function (\ref{fan2}), one needs to integrate the expressions that contains the hypergeometric function $_3F_2$ . To do this one can represent this function  as a integral of a Gauss $_2F_1$ function. It can be done e. g. using formula 2.21.1.4 of Ref.~\cite{PrudnikovBrychkovMarichev}. Then changing the order of integration for $I(1,1,1,2,0,2)$ and using formula 2.21.1.15  and 2.22.2.1 of Ref.~\cite{PrudnikovBrychkovMarichev} one obtains the result in the form:
 \begin{eqnarray}
I(1,1,1,2,0,2)&{=}&r^{\frac{3a-d}{2}{-}4}\left(\frac{S_d}{2}\right)^2
\Bigg\{\frac{\Gamma(\frac{2a-d}{2})\Gamma(2-\frac{2a-d}{2})\Gamma(\frac{d}{2})\Gamma(a-1)\Gamma(2-\frac{a}{2})}{\Gamma(\frac{d+a}{2}-1)}
\times\nonumber\\&&\Bigg[1{+}\frac{\frac{a-d}{2}}{a-2}\Bigg(1+\frac{\frac{a}{2}}{(1-\frac{2a-d}{2})}
\Big(1+\frac{(2-\frac{a}{2})(1-\frac{a-d}{2})}{\frac{a}{2}{(a-3)}}\Big)\Bigg)\Bigg]+\nonumber\\&&
\frac{\Gamma(\frac{2a-d}{2})\Gamma(1-\frac{2a-d}{2})\Gamma(\frac{d}{2})\Gamma(3-a)\Gamma(1-\frac{2a-d}{2})
\Gamma(\frac{a}{2}-2)}{\Gamma(1-\frac{a}{2})\Gamma(\frac{d-a}{2})\Gamma(2-\frac{a-d}{2})}{\times}
\nonumber\\&&\Bigg[(1-\frac{2a-d}{2}){_3F_2\left[\left.\begin{array}{c}
2,3-a,2-\frac{2a-d}{2}\\
2-\frac{a-d}{2},3-\frac{a}{2}\end{array}\right|1\right]}-\nonumber\\&&(\frac{a}{2}-2){_3F_2\left[\left.\begin{array}{c}
2,3-a,2-\frac{2a-d}{2}\\
2-\frac{a-d}{2},2-\frac{a}{2}\end{array}\right|1\right]}\Bigg]+\nonumber\\&&
\frac{\Gamma(\frac{3a-d}{2})\Gamma(2{-}\frac{3a-d}{2})\Gamma(\frac{d}{2})\Gamma^2(\frac{a}{2})\Gamma({}\frac{d-2a}{2})
}{\Gamma^2(\frac{d-a}{2})\Gamma(a)}\Bigg[{_3F_2\left[\left.\begin{array}{c}
2,\frac{a}{2},1+\frac{a-d}{2}\\
a,1+\frac{2a-d}{2}\end{array}\right|1\right]}+\nonumber\\&&\nonumber\\&&\frac{1+\frac{a-d}{2}}{(1+\frac{2a-d}{2})(\frac{3a-d}{2}-1)}{_3F_2\left[\left.\begin{array}{c}
3,1+\frac{a}{2},2+\frac{a-d}{2}\\
a,1+\frac{2a-d}{2}\end{array}\right|1\right]}\Bigg]+\nonumber\\&&
\frac{\Gamma(\frac{d}{2})\Gamma(\frac{a}{2})\Gamma(\frac{d-2a}{2})\Gamma(1{+}\frac{2a-d}{2})\Gamma(\frac{3a-d}{2}{-}2)\Gamma(4{-}\frac{3a-d}{2})
\Gamma(2{-}\frac{2a-d}{2})\Gamma({3{-}a})
}{\Gamma^2(\frac{d-a}{2})\Gamma(1+\frac{a-d}{2})\Gamma(2-\frac{a-d}{2})\Gamma(3-\frac{a}{2})}{\times}\nonumber\\&&\Bigg[{_3F_2\left[\left.\begin{array}{c}
2,3-{a},2-\frac{2a-d}{2}\\
2-\frac{a-d}{2},3-\frac{a}{2}\end{array}\right|1\right]}{+}\nonumber\\&&
\frac{2(2{-}\frac{2a-d}{2})(3{-}a)}{(3{-}\frac{3a-d}{2})(3{-}\frac{a}{2})(2{-}\frac{a-d}{2})}{_3F_2\left[\left.\begin{array}{c}
3,4-{a},3-\frac{2a-d}{2}\\
3-\frac{a-d}{2},4-\frac{a}{2}\end{array}\right|1\right]}\Bigg]\Bigg\}.
\end{eqnarray}

Result of calculation of the last integral $I(1{,}1{,}1{,}2{,}1{,}1)$ which includes  integration of  $f_4(\alpha{,}1{,}1{,}q_1)$   is rather lengthy, therefore we present it in Appendix \ref{ApC}.

Obtained results can be also used within the in minimal subtraction RG scheme \cite{rgbooks}, extracting poles in $\epsilon=4-d$ and $\delta=4-a$. In this case we need to know the first terms in the double expansions in $\epsilon$ and $\delta$ for the corresponding hypergeometric functions.  In the last decade a lot of efforts was put to derive $\epsilon$-expansions for generalised hypergeometric functions analytically \cite{Davydchev00,Davydchev00a,DavydchevKalmykov,Kalmykov06,KalmykovWardYost07,KalmykovWardYost07a,KalmykovKniehl,YostBytevKalmykov,BytevKalmykovKniehl12,GreynatSesma}.
Also algorithms  for the same goal \cite{MochUwerWeinzierl,Weinzierl} exploiting technique of nested sum were elaborated  ready to use within a symbolic computer algebra
system . They found their implementation in several different computer packages \cite{Weinzierl02,MochUwer,HuberMaitre06,HuberMaitre08,BytevKalmykovKniehl11}. To be applied in our case all these methods should be extended to the case of double expansions.

\section{Conclusions}\label{V}
In the present paper we have considered integrals appearing at the renormalization group analysis of the field-theoretical model including long-range correlated disorder. These integrals  depend on the space dimension $d$ and correlation parameter $a$.  They were known analytically only in one-loop approximation at
$d=3$ as functions of $a$.  Integrals of two-loop order were calculated only numerically for $d=3$ and some set of values $a$ in the range between 2 and 3 \cite{PrudnikovPrudnikovFedorenko99,PrudnikovPrudnikovFedorenko00}. Here, we have evaluated two-loop integrals with untrivial massless propagators created by introduction of long-range correlated disorder for vertex function $\Gamma^{(4)}$ in form of combinations of hypergeometric functions with parameters  dependent  on $d$ and $a$.  The  Mellin-Barnes method for evaluation of massive integrals was exploited  at the calculations  together with use of known results \cite{BoosDavydychev}.

Main result of this paper is evaluation of one loop functions $f_2$ (\ref{fan2}) and $f_3$ (\ref{fan3}) for any parameters $d$ and $a$ in form of expressions with hypergeometric functions of one argument (\ref{ff2}), (\ref{f3fin1}). With the help of these expressions (\ref{f1}), (\ref{ff4}) as well as with expressions for functions $f_1$ (\ref{fan1}) and $f_4$ (\ref{fan4})  two-loop integrals with untrivial massless propagators (\ref{int1})-(\ref{int9}) were evaluated.  Combining obtained results with double expansions in $\epsilon$ and $\delta$  of hypergeometric functions  the poles  in $\epsilon$ and $\delta$ can be found for these integrals.
Therefore obtained results can be used at the renormalization-group study of different models of statistical physics that include defects having long range correlations.  Our result also can be exploited  for establishment of relationships  between hypergeometric functions as it was done comparing results for Feynman diagrams evaluated by different methods \cite{KnielTarasov}.

%%%%%%%%%%%%%%%%%%%%%%%%%%%%%%%%%%%%%%%%%%%%%%%%%%%%%%%%%%%%%%%%%%%%%%%%%%%%%%%%%%%%%%%%%%%%%%%%%55
\acknowledgments The author is grateful to Yurij Holovatch for attraction of attention to this subject  and permanent support. I wish to thank Mykola Shpot for
interest in this work, many discussions and critical remarks. This work was supported in part by the 7th FP, IRSES project N269139 ``Dynamics
and Cooperative phenomena in complex physical and biological environments".

\appendix

\section{Some definitions} \label{ApA}

In this appendix we present definitions and some properties of functions used in the paper.

{\bf Gamma function} is an extension of factorial function to noninteger and complex variables. It can be defined by the following expression (see e. g. 1.1.1 in \cite{Erdeley} ):
\begin{equation}
\Gamma(z)=\int_0^{\infty} t^{z-1} {\rm e}^{-t} dt,\qquad\quad z>0.
\end{equation}
Function $\Gamma(z)$ can be analytically continued for negative arguments, where it has series of poles for $z=-n$ at integer $n\ge 0$.
For our calculation two properties of this function are useful. One from them is the duplication formula (see e. g. 1.2.15 in \cite{Erdeley} ):
\begin{equation}\label{dupl}
\Gamma(z)\Gamma\left(z+\frac{1}{2}\right)=2^{1-2z}\sqrt{\pi}\Gamma(2z).
\end{equation}
When the argument of the $\Gamma$ function can be presented as $z-n$, where $n$ is some natural number, the following formula can be used (see e. g. 1.2.3 in \cite{Erdeley} ):
\begin{equation}\label{minn}
\Gamma(z-n)=(-1)^n\frac{\Gamma(z)\Gamma(1-z)}{\Gamma(1-z+n)}.
\end{equation}

Definition of the {\bf generalized hypergeometric series} of one variable is the following~\cite{Erdeley}:
\begin{equation}\label{ghyper}
_pF_q\left[\left.\begin{array}{cccc}
a_1,&a_2,&\dots,&a_p\\
b_1,&b_2,&\dots,&b_q\end{array}\right|z\right]=\sum_{n\ge 0}\frac{ (a_1)_n(a_2)_n\dots (a_p )_n}{n! (b_1)_n(b_2)_n\dots (b_q)_n}z^n,
\end{equation}
where $(a)_n=\frac{\Gamma(a+n)}{\Gamma(n)}$ is {\bf Pochhammer symbol}. Parameters $a_i$ in (\ref{ghyper}) are called {\em numerator parameters} while $b_i$  are called {\em denominator parameters}.

 {\bf Gauss hypergeometric function} can be defined as a particular case of (\ref{ghyper}) at $p=2$ and $q=1$~\cite{Erdeley}:
\begin{equation}\label{2f1}
{_2F_1}\left[\left.\begin{array}{cc}
a,&b\\
c&\end{array}\right|z\right]={_2F_1}[a,b,c;z]=\sum_{n\ge 0}\frac{ (a)_n(b)_n}{n! (c)_n}z^n.
\end{equation}
Function $_2F_1$ has the following integral representation:
\begin{equation}\label{2f1int}
_2F_1[a,b,c;z]=\frac{\Gamma(c)}{\Gamma(b)\Gamma(c-b)}\int_{0}^{1} \frac{x^{b-1}(1-x)^{c-b-1}}{(1-xz)^a} dx,
\end{equation}
and as a contour integral (Mellin-Barnes representation) in complex plane:
\begin{equation}\label{2f1cont}
_2F_1[a,b,c;z]=\frac{\Gamma(c)}{\Gamma(a)\Gamma(b)}\frac{1}{2\pi i}\int^{i\infty}_{-i\infty}ds(-z)^s \Gamma(-s)\frac{\Gamma(a+s)\Gamma(b+s)}{\Gamma(c+s)}.
\end{equation}

Extension of the theory of hypergeometric functions for the case of two variables gives in general case  the {\bf Kamp\'e de F\'eriet functions}~\cite{Exton}:
\begin{eqnarray}\label{kampe}
&&F^{A,B,B'}_{C,D,D'}\left[\left.\begin{array}{c}
a_1,\dots, a_A;b_1,\dots,b_B;{b'}_1,\dots,{b'}_{B'}\\
c_1,\dots, c_C;d_1,\dots,d_D;{d'}_1,\dots,{d'}_{D'}\end{array}\right|x,y\right]=\nonumber\\
\sum_{n\ge 0}\sum_{m\ge 0}&&\frac{ (a_1)_{m{+}n}{\dots}(a_A)_{m{+}n}(b_1)_{m}{\dots}(b_B)_{m}({b'}_1)_{n}{\dots}({b'}_{B'})_{n}}{ (c_1)_{m{+}n}{\dots}(c_C)_{m{+}n}(b_1)_{m}{\dots}(d_D)_{m}({d'}_1)_{n}{\dots}({d'}_{D'})_{n}}\frac{x^n}{n!}\frac{y^m}{m!}.
\end{eqnarray}

Kamp\'e de F\'eriet function $F^{2,0,0}_{0,1,1}$ corresponds to well known {\bf Appell function of fourth type $F_4$}~\cite{Erdeley}:
\begin{equation}\label{appell}
F_4[a,b;c,c'|x,y]=\sum_{n\ge 0}\sum_{m\ge 0}\frac{(a)_{m+n}(b)_{m+n}}{(c)_n (c')_m}\frac{x^n}{n!}\frac{y^m}{m!},
\end{equation}
with the representation via Mellin-Barnes integral:
\begin{eqnarray}
\!\!\!\!\!\!\!F_4[a,b;c,c'|x,y]&{=}&\frac{\Gamma(c)\Gamma(c')}{\Gamma(a)\Gamma(b)}\times\nonumber\\&&
\frac{1}{(2\pi i)^2}\!\int^{i\infty}_{-i\infty}\!\!\!ds\int^{i\infty}_{-i\infty}\!\!\!dt({-}x)^s({}-y)^t \Gamma({-}s)\Gamma({}-t)\frac{\Gamma(a{+}s{+}t)\Gamma(b{+}s{+}t)}{\Gamma(c{+}s)\Gamma(c'{+}t)}.
\end{eqnarray}

\section{Mellin-Barnes method for  massive integrals}\label{ApB}

In this appendix we apply the Mellin-Barnes transform to evaluate  massive integrals (according to Ref.~\cite{BoosDavydychev}).
The main idea consists in a usage of Mellin-Barnes representation of a function  $1/(A+z)^\alpha$ via contour integral in the complex plane:
\begin{equation}
\label{MB}
\frac{1}{(A+z)^\alpha}{=}\frac{1}{2\pi i \Gamma(\alpha)}\int_{-i\infty}^{i\infty} ds \frac{(z)^s} {A^{\alpha+s}}
\Gamma(-s)\Gamma(\alpha+s),
\end{equation}
where  contour in the complex plane of $s$ can be chosen to  separate the ``left" series of poles of the $\Gamma$
functions in the integrand from the ``right" poles. Therefore for evaluation  of integral in (\ref{MB}) the residue
theorem can be used, where contour at infinity  can be closed either in the right  half-plane or in the  left one in order to make the
integrand decrease.
 Let us apply formula (\ref{MB}) to the massive integral of function (\ref{fan1}):
 \begin{equation}
f_1(\alpha,\beta,q_1)=\int_{q_2}\frac{1}{(q^2_2)^\alpha(r+({\bf q}_1+{\bf q}_2)^2)^\beta}.
\end{equation}
It gives us the following result:
\begin{equation}\label{f1_0}
f_1(\alpha,\beta,q_1)=\frac{1}{\Gamma(\beta)2\pi i}\int_{-i\infty}^{i\infty} ds (r)^s\Gamma(-s)\Gamma(\beta+s)\int_{{q}_2}\frac{1}{({q}_2^2)^{\alpha}(({\bf q}_1+{\bf q}_2)^2)^{\beta+s}},
\end{equation}
where the last integration  over $q_2$ is performed only for massless propagators. It can be done with the help of Feynman parametrisation (for more details about Feynman parameters see e. g. \cite{Smirnov,rgbooks}):
\begin{equation}\label{fenpam}
\frac{1}{A^a B^b}=\frac{\Gamma(a+b)}{\Gamma(a)\Gamma(b)}\int_0^1 dx \frac{x^{a-1}(1-x)^{b-1}}{(xA+(1-x)B)^{a+b}}.
\end{equation}
Then integral over $q_2$ in (\ref{f1_0}) can be presented in the following form:
\begin{equation}
\int_{{q}_2}\frac{1}{({q}_2^2)^{a}(({\bf q}_2+{\bf q}_2)^2)^b}=
\frac{\Gamma(a+b)}{\Gamma(a)\Gamma(b)}\int_0^1 dx \int_{{q}_2}\frac{x^{b-1}(1-x)^{a-1}}{(({\bf q}_2+x{\bf q}_1)^2+x(1-x)q_1^2)^{a+b}}.
\end{equation}
Here one can use substitution ${\bf q}_2+x{\bf q}_1\to {{\bf q}'}_2$. Then, in the integration over ${q'_2}$ the angular part can be evaluated separately (see (\ref{spher})), for the rest formula 3.252.11 from \cite{GradshteynRyzhik} can be used. Finally, after performing integration over $x$ one gets result in form:
\begin{equation}\label{q_int}
\int_{{q}_2}\frac{1}{({q}_2^2)^{a}((\vec{q}_1+\vec{q}_2)^2)^b}=\frac{S_d}{2}
\frac{\Gamma({d}/{2})\Gamma(\frac{d}{2}-a)\Gamma({d}/{2}-b)\Gamma(a+b-\frac{d}{2})}{\Gamma(a)\Gamma(b)\Gamma(d-a-b)}(q^2_1)^{{d}/{2}-a-b},
\end{equation}
where  $S_d=\frac{1}{2^{d-1} \pi^{d/2}\Gamma(d/2)}$. Substituting (\ref{q_int}) into (\ref{f1_0}) one has
\begin{eqnarray}
f_1(\alpha,\beta,{q}_1)&=&(q^2_1)^{d/2-\alpha-\beta}\frac{S_d \Gamma(d/2)\Gamma(d/2-\alpha)}{2\Gamma(\alpha)\Gamma(\beta)}
\nonumber\\&&\frac{1}{2\pi i}\int_{-i\infty}^{i\infty} ds
\frac{\Gamma(-s)\Gamma(d/2- \beta-s)\Gamma(\alpha+\beta+s-d/2)}{\Gamma(d-\alpha-\beta-s)}\left(\frac{r}{q^2_1}\right)^s.
\end{eqnarray}
Here, the following change of variables
$s \to d/2-\alpha-\beta-s$ is useful. It transforms the integral to the form
\begin{eqnarray}
f_1(\alpha,\beta,{q}_1)&=&(r)^{d/2-\alpha-\beta}\frac{S_d \Gamma(d/2) \Gamma(d/2-\alpha)}{2\Gamma(\alpha)\Gamma(\beta)}
\nonumber\\&&\frac{1}{2\pi i}\int_{-i\infty}^{i\infty} ds
\frac{\Gamma(-s)\Gamma(\alpha+s)\Gamma(\alpha+\beta+s-d/2)}{\Gamma(d/2+s)}\left(\frac{q^2_1}{r}\right)^s.
\end{eqnarray}
Using definition (\ref{2f1cont}) one obtains
\begin{equation}
f_1(\alpha,\beta,{q}_1)=(r)^{d/2-\alpha-\beta}
\frac{S_d \Gamma(\alpha+\beta-d/2) \Gamma(d/2-\alpha)}{2\Gamma(\beta)}\,
{_2F_1\!\left[\left.\begin{array}{c}
\alpha,\alpha+\beta-d/2\\ d/2\end{array}\right|-\frac{q^2_1}{r}\right]}.
\end{equation}

\section{Expression for integral $I(1,1,1,2,1,1)$} \label{ApC}
Here, we present a way for calculation integral $I(1,1,1,2,1,1)$. It can be done according to the formula (\ref{int111211}). Also it is obvious that
\begin{equation}\label{dif}
I(1,1,1,2,1,1)=-\frac{1}{3}\frac{\partial}{\partial r}I(1,1,1,1,1,1),
\end{equation}
where the integral in the r.h.s. of (\ref{dif}) also is cal\-culated via the same func\-tion $f_4(\!\frac{d{-}a}{2},1,1,q_1^2)$:
 \begin{equation}\label{int111111}
I(1,1,1,1,1,1)=\int_{q_1}d q_1\frac{(q_1^2)^{\frac{a-d}{2}}}{(r+q^2_1)}f_4\left(\frac{d-a}{2},1,1,q_1^2\right).
\end{equation}

Then we can substitute to (\ref{int111111}) the expression (\ref{ff4}) and use Mellin-Barnes representation of the corresponding Kamp\'e de F\'eriet functions in the expression for $f_4(\alpha,1,1,q_1)$. Obtained expression is quite long and cumbersome and we present it in the following form:

\begin{eqnarray}
I(1,1,1,1,1,1)&{=}&r^{\frac{3a-d}{2}{-}4}\left(\frac{S_d}{2}\right)^2\Gamma\left(\frac{d}{2}\right)\left\{\frac{}{}A(a,d){+}B(a,d){+}\right.\nonumber\\
&&(\frac{d}{2}{-}1) \frac{\Gamma \left(2{-}a{-}\frac{d}{2} \right) \Gamma \left({-}1{+}a{-}\frac{d}{2}\right)}{\Gamma \left(\frac{d}{2}\right)} \left[\Gamma \left(1-\frac{a}{2}\right) \Gamma
   \left(\frac{a}{2} \right) {_2F_1\left[\left.\begin{array}{c}
1,2-\frac{d}{2}\\ \frac{3}{2}\end{array}\right|\frac{1}{4}\right]}\right.
   +\nonumber\\&& \left. 2^{a{-}2}\left.\frac{  \Gamma(\frac{3}{2})\Gamma (3{-}\frac{a{+}d}{2}) \Gamma \left(2{-}\frac{a}{2}\right)
   \Gamma \left(\frac{a}{2} {-}1\right) }{\Gamma
   \left(2{-}\frac{d}{2}\right) \Gamma \left(\frac{5-a}{2}\right)}{_2F_1\left[\left.\begin{array}{c}
1,3{-}\frac{a+d}{2}\\ \frac{5-a}{2} \end{array}\right|\frac{1}{4}\right]}\right]\right\},
\end{eqnarray}
where $A(a,d)$, $B(a,d)$ are results of integration of the second and the third terms respectively:
\begin{eqnarray}
A(a,d)&=&
\frac{ \Gamma \left(1{+}\frac{2a{-}d}{2}\right)\Gamma \left(\frac{3a{-}d}{2}\right)\Gamma \left(\frac{d{-}2a}{2}\right)\Gamma \left(1{-}\frac{3a{-}d}{2}\right) \Gamma^2 \left(\frac{a}{2}\right)}{ \Gamma^2 \left(\frac{d-a}{2}\right)\Gamma \left(a\right)\Gamma \left(1{+}a{-}\frac{d}{2}\right)}{\times}\nonumber\\&&{F^{0,3,3}_{2,0,0}\left[\!\!\left.\begin{array}{c}
0;1,\frac{a}{2},1{+}\frac{a-d}{2}; 1,\frac{a}{2},1{+}\frac{a-d}{2}\\
a,1{+}a{-}\frac{d}{2};0;0\end{array}\!\!\right|1,1\right]}{+}\nonumber\\&&
\frac{ \Gamma \left(1{+}a{-}\frac{d}{2}\right)\Gamma \left(\frac{3a{-}d}{2}{-}1\right)\Gamma \left({-}a{+}\frac{d}{2}\right)\Gamma \left(2{-}\frac{3a{-}d}{2}\right) \Gamma \left(\frac{a}{2}\right)\Gamma \left(2{-}{a}\right)\Gamma \left(1{-}a{+}\frac{d}{2}\right)}{ \Gamma^2 \left(\frac{d-a}{2}\right)\Gamma \left(2-\frac{a}{2}\right)\Gamma \left(1{+}\frac{a-d}{2}\right)\Gamma \left(1{-}\frac{a-d}{2}\right)}{\times}\nonumber\\&&{F^{0,3,3}_{2,0,0}\left[\!\!\left.\begin{array}{c}
0;1,\frac{a}{2},1{+}\frac{a-d}{2}; 1,2{-}{a},1{-}a{+}\frac{d}{2}\\
2{-}\frac{a}{2},1{-}\frac{a-d}{2};0;0\end{array}\!\!\right|1,1\right]}{+}\nonumber\\&&
\frac{ \Gamma \left(\frac{a}{2}{-}1\right)\Gamma \left(\frac{a{-}d}{2}\right)\Gamma \left(1{+}\frac{2a{-}d}{2}\right)\Gamma \left(\frac{3a{-}d}{2}{-}2\right) \Gamma \left({-}a{+}\frac{d}{2}\right)\Gamma \left(2{-}\frac{a{-}d}{2}\right)\Gamma\left(3{-}\frac{3a{-}d}{2}\right) }{ \Gamma^2 \left(1{+}\frac{a-d}{2}\right)\Gamma^2 \left(\frac{d-a}{2}\right)\Gamma^2 \left(1{-}\frac{a-d}{2}\right)\Gamma \left(2-\frac{a}{2}\right)}{\times}\nonumber\\&&\Gamma \left(3{-}a\right)\Gamma \left(2{-}\frac{2a{-}d}{2}\right){F^{2,2,1}_{0,3,2}\left[\!\!\left.\begin{array}{c}
3{-}a,2{-}a{+}\frac{d}{2};1,3{-}\frac{3a-d}{2}; 1\\
0;2{-}\frac{a}{2},1{-}\frac{a-d}{2},3{+}\frac{3a-5d}{2};2{-}\frac{a}{2},1{-}\frac{a-d}{2}\end{array}\!\!\right|1,1\right]}{+}\nonumber\\&&
\frac{1}{2}\Gamma \left(1{-}\frac{a}{2}\right)\Gamma \left(\frac{a}{2}\right)\Gamma \left(2{-}\frac{d}{2}\right){_2F_1}\left[\left.\begin{array}{c}
1,2-\frac{d}{2}\\3/2\end{array}\right|\frac{1}{4}\right]{+}\nonumber\\&&
\frac{ \Gamma \left({a}{-}2\right)\Gamma \left(1{+}a{-}\frac{d}{2}\right)\Gamma \left({-}a{+}\frac{d}{2}\right)\Gamma \left(3{-}a\right) \Gamma \left(2{-}\frac{a}{2}\right)\Gamma \left(3{-}\frac{a{+}d}{2}\right)}{ \Gamma \left(1{+}\frac{a-d}{2}\right)\Gamma \left(\frac{d-a}{2}\right)\Gamma \left(1{-}\frac{a-d}{2}\right)\Gamma \left(2-\frac{a}{2}\right)}{\times}\nonumber\\&&{F^{2,1,1}_{0,2,2}\left[\!\!\left.\begin{array}{c}
2{-}a,3{-}\frac{a+d}{2};3{-}a; 1\\
0;3{+}a{-}{2d},2{-}\frac{d}{2};2{-}\frac{a}{2},1{-}\frac{a-d}{2}\end{array}\!\!\right|1,1\right]}{+}\nonumber\\&&
\frac{ \Gamma \left({2}{-}a\right)\Gamma \left(1{+}a{-}\frac{d}{2}\right)\Gamma \left(\frac{3a-d}{2}{-}1\right)\Gamma \left({-}a{+}\frac{d}{2}\right) \Gamma \left(1{-}a{+}\frac{d}{2}\right) \Gamma \left(\frac{a}{2}\right)\Gamma \left(2{-}\frac{3a{-}d}{2}\right)}{ \Gamma \left(1{+}\frac{a-d}{2}\right)\Gamma^2 \left(\frac{d-a}{2}\right)\Gamma \left(1{-}\frac{a-d}{2}\right)\Gamma \left(2-\frac{a}{2}\right)}{\times}\nonumber\\&&{F^{2,1,1}_{0,2,2}\left[\!\!\left.\begin{array}{c}
\frac{a}{2},1{+}\frac{a-d}{2};1; 1\\
0;a{-}{1},a{-}\frac{d}{2};2{-}\frac{a}{2},1{-}\frac{a-d}{2}\end{array}\!\!\right|1,1\right]},
\end{eqnarray}

\begin{eqnarray}B(a,d)&=&
\frac{ \Gamma \left(a{-}\frac{d}{2}\right) \Gamma \left(1{-}\frac{2a-d}{2}\right) \Gamma \left(\frac{a}{2}\right) \Gamma \left(1{-}\frac{a}{2}\right)}{ \Gamma \left(\frac{d}{2}\right)}{F^{0,3,3}_{2,0,0}\left[\left.\begin{array}{c}
0;1,\frac{a}{2},1{+}\frac{a-d}{2}; 1,1{-}\frac{a}{2},\frac{d-a}{2}\\
1,\frac{d}{2};0;0\end{array}\!\!\right|1,1\right]}{+}\nonumber\\&&
\frac{ \Gamma \left(\frac{a}{2}{-}1\right)\Gamma \left(a{-}\frac{d}{2}\right) \Gamma^2 \left(1{-}a{+}\frac{d}{2}\right) \Gamma \left(2{-}a\right) \Gamma \left(1-\frac{a}{2}\right)}{ \Gamma \left(1-\frac{a}{2}\right)\Gamma \left(\frac{d-a}{2}\right)\Gamma \left(1{-}\frac{a{-}d}{2}\right)}{\times}\nonumber\\&&{F^{0,3,3}_{2,0,0}\left[\!\!\left.\begin{array}{c}
0;1,2{-}a,1{-}a{+}\frac{d}{2}; 1,\frac{a}{2},1{+}\frac{a-d}{2}\\
2{-}\frac{a}{2},1{-}\frac{a-d}{2};0;0\end{array}\!\!\right|1,1\right]}{+}\nonumber\\&&
\frac{ \Gamma \left(3{-}\frac{a}{2}\right)\Gamma \left(\frac{a}{2}{-}2\right)\Gamma \left(-\frac{a}{2}\right)\Gamma \left(a{-}\frac{d}{2}\right) \Gamma \left(1{-}a{+}\frac{d}{2}\right)\Gamma\left(\frac{d-a}{2}{-}1\right)\Gamma\left(3{-}\frac{d}{2}\right) }{ \Gamma \left(1-\frac{a}{2}\right)\Gamma \left(\frac{d-a}{2}\right)\Gamma \left(1{-}\frac{a{-}d}{2}\right)}{\times}\nonumber\\&&{F^{2,1,1}_{0,2,2}\left[\!\!\left.\begin{array}{c}
2,3{-}\frac{d}{2};1; 1\\
0;1{+}\frac{a}{2},2{+}\frac{a-d}{2};2{-}\frac{a}{2},1{-}\frac{a-d}{2}\end{array}\!\!\right|1,1\right]}{+}\nonumber\\&&
\frac{ \Gamma \left(2{-}\frac{d}{2}\right)\Gamma \left({-}1{+}a{-}\frac{d}{2}\right)\Gamma \left(a{-}\frac{d}{2}\right)\Gamma \left(1{-}a{+}\frac{d}{2}\right) \Gamma \left(2{-}a{+}\frac{d}{2}\right)}{ 2\Gamma \left(1-\frac{a}{2}\right)\Gamma \left(\frac{a}{2}\right)}{_2F_1}\left[\left.\begin{array}{c}
1,2-\frac{d}{2}\\3/2\end{array}\right|\frac{1}{4}\right]{+}\nonumber\\&&
\frac{ \Gamma \left(3{-}a\right)\Gamma \left(a{-}2\right)\Gamma \left(a{-}\frac{d}{2}\right)\Gamma \left(\frac{d}{2}-1\right) \Gamma \left(1{-}a{+}\frac{d}{2}\right)\Gamma \left(2{-}\frac{a}{2}\right)\Gamma \left(3{-}\frac{a+d}{2}\right)}{ \Gamma \left(1{+}\frac{a-d}{2}\right)\Gamma \left(\frac{d-a}{2}\right)\Gamma \left(2{-}\frac{a}{2}\right)\Gamma \left(1{-}\frac{a-d}{2}\right)}{\times}\nonumber\\&&{F^{2,1,0}_{0,2,1}\left[\!\!\left.\begin{array}{c}
2{-}\frac{a}{2},3{-}\frac{a+d}{2};1; 0\\
0;2{-}\frac{a}{2},1{-}\frac{a-d}{2};2-\frac{d}{2}\end{array}\!\!\right|1,1\right]}{+}\nonumber\\&&
\frac{ \Gamma \left(2{-}a\right)\Gamma \left(a{-}1\right)\Gamma \left(a{-}\frac{d}{2}\right)\Gamma \left(\frac{a}{2}-1\right) \Gamma^2 \left(1{-}a{+}\frac{d}{2}\right)}{ \Gamma \left(1{-}\frac{a}{2}\right)\Gamma \left(\frac{d-a}{2}\right)\Gamma \left(1{-}\frac{a-d}{2}\right)\Gamma \left(a{-}1\right)}{\times}\nonumber\\&&{F^{2,1,1}_{0,2,2}\left[\!\!\left.\begin{array}{c}
\frac{a}{2},1{+}\frac{a-d}{2};1; 1\\
0;a{-}1,a{-}\frac{d}{2};2{-}\frac{a}{2},1{-}\frac{a-d}{2}\end{array}\!\!\right|1,1\right]}
\end{eqnarray}.

\end{document}